\documentclass{aa}
\usepackage{graphicx,psfig}

\newcommand{\Msun} {M$_\odot$}
\newcommand{\Lsun} {L$_\odot$}
\newcommand{\Rsun} {R$_\odot$}

\newcommand{\Tstar} {T$_{\rm{eff}}$}
\newcommand{\Lstar} {L$_\star$}
\newcommand{\Mstar} {M$_\star$}
\newcommand{\Rstar} {R$_\star$}

\newcommand{\um} {$\mu$m}

\newcommand{\Myr} {M$_\odot$/yr}
\newcommand{\kms} {km/s}
\newcommand{\Ha} {H$\alpha$}
\newcommand{\Roph} {$\rho$~Oph}
\newcommand{\Pab} {Pa$\beta$}
\newcommand{\Brg} {Br$\gamma$}
\newcommand{\MJ} {M$_J$}
\newcommand{\Macc} {$\dot M_{ac}$}
\newcommand{\Lacc} {L$_{ac}$}

\newcommand{\simless}{\mathbin{\lower 3pt\hbox
      {$\rlap{\raise 5pt\hbox{$\char'074$}}\mathchar"7218$}}} 
\newcommand{\simgreat}{\mathbin{\lower 3pt\hbox
     {$\rlap{\raise 5pt\hbox{$\char'076$}}\mathchar"7218$}}} 

\begin{document}

\title{Accretion in Brown Dwarfs: an Infrared View
\thanks{
Based on observations collected at the European Southern Observatory, Chile.
}}

\author{
A. Natta\inst{1},
L. Testi\inst{1},
J. Muzerolle \inst{2},
S. Randich \inst{1},
F. Comer\'on \inst{3},
P. Persi \inst{4}
}
\institute{
    Osservatorio Astrofisico di Arcetri, INAF, Largo E.Fermi 5,
    I-50125 Firenze, Italy
\and
Steward Observatory, University of Arizona, 933 North Cherry Avenue, Tucson, AZ 85721, USA
\and
European Southern Observatory, Karl-Schwarzschild-Strasse 2, 85748 Garching, Germany
\and
Istituto Astrofisica Spaziale e Fisica Cosmica, CNR, Via del Fosso del Cavaliere, 00133 Roma, Italy
}

\offprints{natta@arcetri.astro.it}
\date{Received ...; accepted ...}

\authorrunning{Natta et al.}
\titlerunning{BD accretion}

\abstract{ 
This paper presents a study of the accretion properties of
19 very low mass objects (\Mstar$\sim 0.01-0.1$ \Msun)
in the regions Chamaeleon I and \Roph. 
For 8 objects we obtained high resolution \Ha\ profiles and determined 
mass accretion rate \Macc\ { and accretion luminosity \Lacc.}
\Pab\ is detected in emission in 7 of the 10 \Roph\ objects, but only in one in Cha~I.
{ Using objects for which we have both a determination of \Lacc\ from \Ha\
and a \Pab\ detection,} we show that
the correlation between the \Pab\ luminosity
and luminosity \Lacc, found by Muzerolle et al.~(\cite{Mea98})
for T Tauri stars in Taurus, extends  to objects with mass $\sim$ 0.03 \Msun;
L(\Pab) can be used to measure \Lacc\ also in the substellar regime.
The results were less conclusive for \Brg, which was  detected only in 2 objects, neither of which had an \Ha\ estimate of \Macc.
Using the relation between L(\Pab) and \Lacc\ we determined the accretion rate for all the objects in our sample { (including those with no \Ha\ 
spectrum), }
more than doubling the number of substellar objects with known \Macc. When plotted as a function of the mass of the central object together with data from the literature, 
our results confirm the trend of lower \Macc\ for lower \Mstar, although with a large spread. Some of the spread is probably due to an age effect;
our very young objects in \Roph\ have on average an accretion rate
at least one order of magnitude higher than objects of similar mass in older regions.
As a side product, we found that the width of \Ha\ measured at 10\% peak intensity is not only a qualitative indicator of the accreting nature of very low mass objects, but can be used to obtain a quantitative,
although not very accurate, estimate of \Macc\ over a large mass range,
from T Tauri stars to brown dwarfs.
Finally, we found that
some of our objects show evidence of mass-loss in their optical spectra.
}

\maketitle

\section{Introduction}

Several objects of very low mass  discovered in regions of star formation
have infrared excess typical of circumstellar disks (e.g., Natta and Testi \cite{NT01}, Testi et al.~\cite{Tea02}; Natta et al.~\cite{Nea02})
and show signs of 
accretion-related activity (Jayawardana et al.~\cite{Jay02};
White and Basri \cite{WB03}; Muzerolle et al.~\cite{Mea03}):  they undergo an evolutionary phase similar to that
of classical T Tauri stars (TTS) (Jayawardana et al.~\cite{Jay03b}). This evidence sets  
potentially important constraints on the formation of very low mass objects,
and on the dominance of dynamical processes in star formation in general (e.g.,
Bate et al.~\cite{BBB03}),
which, however,  need to be made more quantitative.
Statistics of disks, measurements of their masses, characterization of the
accretion as a function of the mass of the central objects in regions of
different age and properties are required, and are being accumulated,
exploiting fully the capabilities of 8-m class telescopes.

In this paper we discuss one important parameter in this scenario,
namely the accretion rate in very low mass objects, which, together with the 
disk mass,
determines the timescale for disk survival. Mass accretion rates have been
determined for a handful of  very low mass objects, mostly from
model-fitting of the \Ha\ profiles (Muzerolle et al.~\cite{Mea03}).
The results indicate that the trend of lower accretion rates for
objects of lower mass,  known for TTS 
(White and Ghez \cite{WG01}; Rebull et al.~\cite{Rea00}) continues in the substellar regime, where typical accretion rates are much lower
than in TTS and can be as low as $\sim 5\times 10^{-12}$ \Myr\ (Muzerolle et al.~\cite{Mea00}).  
There is also evidence that, again as for TTS (e.g. Calvet et al.~\cite{CHS00}), the
fraction of accreting objects among very low mass objects is higher in younger regions (Jayawardana et al.~\cite{Jay02}, \cite{Jay03b}). 
All this points  to a continuity in the accretion properties across the range of mass explored so far, which suggests a similar formation mechanism from
solar to sub-stellar masses.

These
results, however, are based on few measurements only, and need to be confirmed.
In particular, it is important to extend them to younger,
more embedded regions, where the use of \Ha\ to derive the accretion rate is
made hard or altogether impossible by the high extinction. In these regions,
it is natural to use instead  infrared hydrogen recombination lines,
such as \Pab\ or \Brg.  For TTS, it has been shown that the
flux in these lines provides a good estimate of the accretion rate
(Muzerolle et al.~\cite{Mea98}). The purpose of this paper is to extend this method to the
sub-stellar regime,  by obtaining near-IR spectroscopy of a sample
of well-known objects in two star-forming regions (Chamaeleon I and
\Roph) for which we have also
acquired and analyzed high-resolution \Ha\ profiles. 

The range of masses of interest goes from
the low end of the stellar range ($\simless 0.2$\Msun) to  brown
dwarfs (BD: $<$0.075 \Msun) and planetary-mass objects, below the
deuterium-burning limit ($<$0.013 \Msun\ or 13 \MJ). In fact,
these distinctions are meaningful only ``a posteriori",  when
bona-fide stars reach the main sequence, not
in young star-forming regions, where
all objects, independently of their mass,
derive their energy from contraction. 
For the sake of simplification
(and in view of the uncertainties affecting mass determinations), we
will talk in the following  of very low mass objects or VLMOs.

In \S 2 we detail the observations, obtained with UVES and ISAAC on VLT.
The results are presented in \S 3 and discussed in \S 4.
A summary follows in \S 5.

\section {Observations}

\subsection {Sample}
\begin{figure}
\begin{center}
\leavevmode
\centerline{ \psfig{file=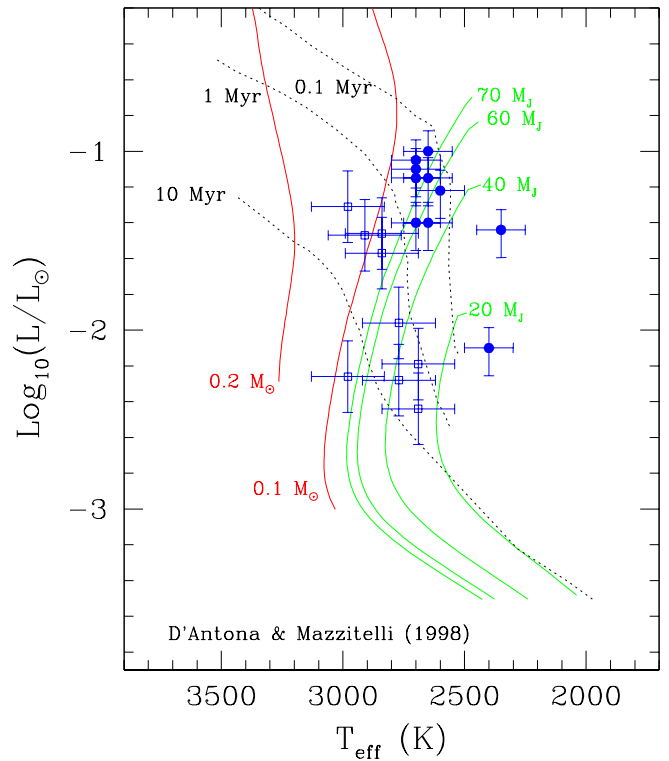,width=9.cm,angle=0} }
\end{center}
\caption{Location of the sample objects in the HR diagram. Filled
dots are objects
in \Roph, open squares objects in Cha~I. The tracks are from D'Antona and Mazzitelli
(\cite{DM97}) and their unpublished 1998 update.
}
\label{HR}
\end{figure}

We have collected a total of 19 objects, 9 in Chamaeleon I and 10 in \Roph.
Their properties are listed in Table~1. For the Cha~I sample, spectral types,
effective temperatures, extinctions, luminosities and masses are from
Comer\'on et al~(\cite{Cea00}).
Quoted uncertainties are $\pm$150~K in \Tstar, $\pm$0.2 dex in Log \Lstar, and a factor { of 2} in \Mstar.
Stellar parameters for the \Roph\ sources are from Natta et al.~(\cite{Nea02}),
with the exception of GY10, not included
in the Natta et al. sample,   for which we adopt the  parameters
of Wilking et al.~(\cite{WGM99}). 
Uncertainties are $\pm$100~K in \Tstar, $\pm$20--30\% in \Lstar, $\pm$ 20--40\% in \Mstar { (see Testi et al.~\cite{Tea02} for a detailed discussion).}
The sample contains a number of bona-fide BDs, several objects that are
either very low mass stars or relatively high-mass BDs, one object
(\Roph -033) which has an estimated  mass of $\sim$10 \MJ\ only.
Based on their location on the HR diagram (see Fig.~\ref{HR}), the VLMOs in Cha~I are significantly
older than those in \Roph\ ($>$2 Myr versus $<$1 Myr, respectively).

\subsection{Low resolution near-infrared spectroscopy}

Near-infrared low resolution spectra in the J and K bands were obtained 
for all our targets using the ISAAC near-infrared camera and spectrograph at the
ESO-VLT UT1 telescope. The observations were carried out partly in Visitor Mode
on April 8, 2002, and partly in Service Mode from March to July 2003
(see Table~2; Table~2 is only available in
electronic form at "http://www.edpsciences.org").
We used the 0.6~arcsec slit and the low resolution grisms 
that offered a $\sim$800 resolution across the wavelength ranges
1.0--1.3~$\mu$m and 1.95--2.45~$\mu$m. The observing sequence was composed by 
a set of pairs of spectra with the target in different positions along the slit,
to allow for an efficient and accurate sky subtraction. The on-source
integration times varied from 25 to 40 minutes depending on target brightness
and observing band (the J-band observations were typically 40--30\% shorter
than the K-band ones). Standard calibrations (flats and lamps)
and telluric standards spectra were obtained for each observation.
Wavelength calibration was performed using the lamps observations and
adjusted by measuring a few OH sky lines before sky removal.
We did not attempt to obtain flux calibrations.

\subsection{High resolution optical spectroscopy}

High resolution optical spectra for the eight brightest (in the optical bands)
objects in our
sample were collected using the UVES spectrograph on the ESO-VLT UT2 telescope
(see Table~2).
For each target a series of three to five $\sim$45~min exposures
were obtained
in the period from March to June 2003.
All spectra were obtained using the standard UVES setup with central 
wavelength 580nm. The chosen setup allowed us to cover  a nominal spectral 
range from $\sim$450 to 680nm. However, due to the very low signal
obtained in the blue part of the spectra (450--580 nm),
only the red part was analyzed. The slit used was 12 arcsec
long and 1 arcsec wide, giving a resolution R$\sim$40000. 
Standard calibrations were obtained by the ESO staff 
as part of the Service Mode and the data were reduced using the UVES pipeline.
All spectra were of excellent quality and the H$\alpha$ line was detected with
high signal to noise in all individual spectra. 

\begin{table*}
\begin{flushleft}
\caption{ Object Properties and Hydrogen Lines}
\begin{tabular}{lcccccccccccc}
\hline\hline
&&&&&&&&&&&&\\
Star  & ST&  \Tstar& Log\Lstar& A$_V$  &  \Mstar& \Ha 10\% &EW(\Ha)$^a$ &  EW(\Pab)$^a$&EW(\Brg)$^a$&  Log$\dot M_{ac}^b$& Log$\dot M_{ac}^c$\\
      &   &   (K)  &(\Lsun)& (mag) & (\MJ)& (\kms )& (\AA) &(\AA)& (\AA)& (\Msun/y)&
(\Msun/y)\\
&&&&&&&&&&&&\\
\hline
Cha \Ha 1& M7.5& 2770& -1.96& 0.2 &40 & 155&35& $<$0.3 &$<$1&$<-$12&$<-$11.4 \\
Cha \Ha 2& M6.5& 2910& -1.47& 0.8 & 70 &366&33 &0.3&$<$1.5 &$-$10.0&$-$10.8\\
Cha \Ha 3& M7& 2840& -1.46& 0.3 & 60   &142&10& $<$0.9 &$<$1.2 &$<-$12&$<-$10.1\\
Cha \Ha 5& M6& 2980& -1.31& 1.0 & 100  & 137&6.5&$<$0.9 &$<$1.7 &$<-$12&$<-$10.0\\
Cha \Ha 6& M7& 2840& -1.57& 0.26 & 50  & 274&48&$<$0.3 &$<$1 & $-$10.5&$<-$10.8\\
Cha \Ha 7& M8& 2690& -2.19& 0.3 & 30   &-- &--&$<$0.6&$<$1.0  &--&$<-$11.1\\
Cha \Ha 9& M6& 2980& -2.26& 1.5 & 70   &-- &--&$<$0.6 &$<$1.0  &--&$<-$11.4\\
Cha \Ha 10&M7.5& 2770& -2.28& 0.1&40   &-- &--&$<$0.3 &$<$0.9 & --&$<-$12.0\\
\medskip
Cha \Ha 11& M8& 2690& -2.44& 0.0 &30   &-- &--&$<$0.3 &$<$0.9 &--&$<-$12.1\\
\Roph -023& M7& 2650& -1.40& 8.0 & 40 & -- &--& 1.80&$<$1&-- &$-$9.3\\
\Roph -030& M6& 2700& -1.15& 3.0 &  60  & 292 &30& 0.30 &$<$1&$-$10.8&$-$10.1\\
\Roph -032& M7.5& 2600& -1.22& 2.0 &40  & 248 &50& 0.40 &$<$0.9 &$-$10.5&$-$9.8\\
\Roph -033& M8.5& 2400&-2.10 & 7.0& 10  & -- &-- &$<$0.7 &$<$2.1  &--&$<-$10.3\\
\Roph -102& M6& 2700& -1.10& 3.0 &  60  & 377 &40& 2.0 &$<$1 &$-$9.0&$-$8.9\\
\Roph -160& M6& 2700& -1.40& 6.0 & 45 & -- &--& 3.3 &4.0 &--&$-$9.0\\
\Roph -164& M6& 2700& -1.05& 6.0 &  60  & --&-- & 0.80&$<$1.6 &--&$-$9.3\\
\Roph -176& M6& 2650& -1.15& 7.0 &  50  & --& --& $<$0.5 &$<$1.3&--&$<-$9.7\\
\Roph -193& M6& 2650& -1.00& 7.5 &  60  & --& --& 1.8 &2.9&--&$-$8.9\\
\Roph -GY10&M8.5 &2350 &-1.44 & 18 &30 & -- &--& $<$0.5&$<$1.0  &--&$<-$9.0\\
\hline
\hline
\label{stars}
\end{tabular}
\end{flushleft}
$^a$ Positive values indicate emission.
\newline
$^b$ Accretion rates from \Ha\ profiles.
\newline
$^c$ Accretion rates from  \Pab\ luminosities.
\end{table*}

\section {Results}

\subsection {Accretion rates from \Ha}

\begin{figure}
\begin{center}
\leavevmode
\centerline{ \psfig{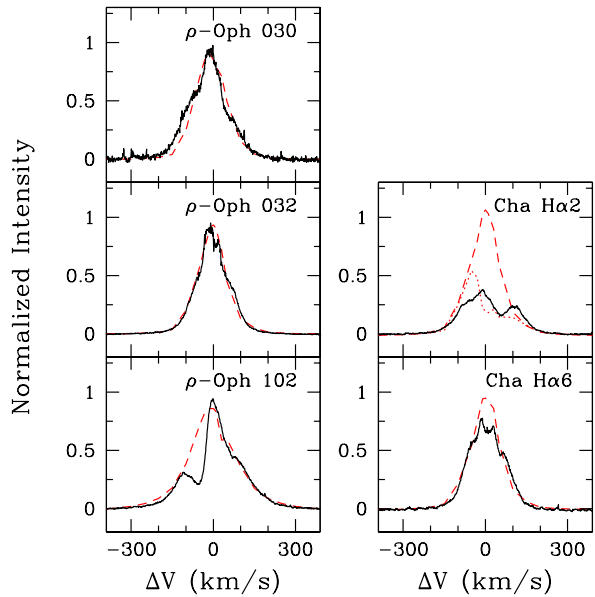} }
\end{center}
\caption{Normalized average \Ha\ profiles for the 5 objects with broad lines.
The solid curves are the observed profiles, the dashed ones are the
model fits. For Cha \Ha 2, the dotted lines shows a model with an
ad-hoc temperature profiles (see text).The stellar continuum has been
subtracted.}
\label{Ha_fits}
\end{figure}

All eight observed objects have \Ha\ emission,
in seven cases with equivalent width $\simgreat 10$ \AA.
Of these, five (2 in Cha~I and 3 in \Roph)
show in all our observations broad
\Ha\ profiles. This, as discussed by White and Basri (\cite{WB03})
 and  Jayawardana et al.~(\cite{Jay02}), is indication of 
accretion-related rather than chromospheric activity.
For each of these objects, we have computed an average profile, shown  in 
Fig.~\ref{Ha_fits},
and compared it to the prediction of the magnetospheric accretion models
developed by  Muzerolle et al.~(\cite{Mea01}) and
recently applied to a sample of sub-stellar objects
by Muzerolle et al.~(\cite{Mea03}). 
Accretion rates
derived from \Ha\ profiles agree well with those
obtained from veiling in TTS (e.g., Muzerolle et al.~\cite{Mea98b}) and
in the few VLMOs for which \Macc\ could be estimated from veiling
measurements (Muzerolle et al.~\cite{Mea03}).

The underlying assumption is  that the entire \Ha\ (as well as the other
hydrogen lines)
comes from accreting matter, and that the
contamination from wind/jet emission is negligible. This is not
necessarily true
in all strongly accreting TTS (see, for example, Bacciotti et al.~\cite{Bac02}),  but is
 probably 
a good approximation for our objects (see \S 3.4).

For all objects
we have adopted a stellar mass \Mstar=50 \MJ, radius 0.5 \Rsun\ and a
magnetospheric truncation radius of 2.2--3 \Rstar.
The results are not very sensitive to the exact values of these
parameters.
The fitting procedure gives for each object a value of \Macc\ { and of the
inclination angle $i$.
The resulting accretion rates are given in Table~\ref{stars}; they range from
$10^{-9}$ \Myr\ for \Roph-102 to $\sim 10^{-11}$ \Myr\ for \Roph-030.
The inclination angle is $\sim 65-75$ deg for Cha \Ha 2, 
\Roph-030, \Roph-032, $\sim 80$ deg for Cha \Ha 6,and $\sim$30 deg for \Roph-102.}
The fits are  of acceptable quality, with the exception of
Cha \Ha 2, whose  broad and flat profile (clearly present in all
the observations) cannot be reproduced by the current models
(similar profiles are observed in few other VLMOs and
discussed in Muzerolle et al.~\cite{Mea03}).
The fit can be somewhat improved by adopting
an ad-hoc temperature profile for the
accretion flow, whereby the temperature has been reduced relative to the
fiducial model in the outer part of the flow near the disk.
This results in less emission near the line center, the most
uncertain part of the profiles given the approximations
used in the radiative transfer calculations (see Muzerolle et al.~\cite{Mea01}).
Nevertheless, both  models for Cha \Ha 2 have {similar inclination}
and an  accretion rate  of the order of $10^{-10}$
\Myr, which we will adopt  in the following.
As discussed by Muzerolle et al.~(\cite{Mea03}), the values of  \Macc\
derived from the \Ha\ profiles
should be accurate within a factor  of $\sim 3-5$.

A second group of 3 objects, all in Cha~I,
(Cha \Ha 1, Cha \Ha 3, Cha \Ha 5)
has always narrow \Ha\ profiles, typical of cromospheric activity. For these
objects, we estimate an upper limit to \Macc\
of about $10^{-12}$ \Myr, for which the magnetospheric accretion models
predict a line emission below the detection limit.

In Fig.~\ref{haw} we compare our estimates of \Macc\ to those
of other VLMOs and TTS from the literature.
The figure plots \Macc\ as a function of
the  full width of \Ha\ measured at 10\% of the
peak intensity (after continuum subtraction).
Dots  are VLMOs (mass $\simless 0.2$ \Msun),
where 
\Macc\ is measured by fitting
the \Ha\ profiles 
(filled dots from this paper, open dots from Muzerolle et al.~\cite{Mea03});
pentagons are VLMOs with values of
\Macc\ high enough to produce measurable veiling in the optical
(White and Basri \cite{WB03}; Muzerolle et al.~\cite{Mea00}, \cite{Mea03}; 
Barrado y Navascu\'es et al.~\cite{BMJ04}).
{ The squares  show the location on this diagram of a sample of more
massive TTS, from which \Macc\ has been derived from veiling
(Gullbring al.~\cite{Gea98}), and the 10\% \Ha\ width are from
high resolution \Ha\ profiles obtained by
Edwards et al.~(\cite{Eea94}).}
The total sample spans a range of masses from about 0.04 to 0.8 \Msun.
Fig.~\ref{haw} shows that
our determinations of \Macc\ agree with the
general trend shown by other objects from the literature, and that there is 
continuity
between measurements obtained from
veiling (which are fully independent of the \Ha\ width)
and those derived from \Ha\ profiles.


In the sub-stellar regime, the \Ha 10\% width,
which can be derived directly from the observations,
is considered a good indicator of accretion,
with the separation between accretors and non-accretors set at 
\Ha 10\% width between $\sim$200 \kms\ (Jayawardana et al.~\cite{Jay03b})
and $\sim$ 270 \kms\ (White and Basri \cite{WB03}).
Fig.~\ref{haw} confirms this result. 
Adopting $\sim$200 \kms\ as a limit, one would missclassify one accreting
object out of 23
(MHO-4, with \Macc$\sim 1.5\times 10^{-11}$ \Myr\ and \Ha\ 10\% width
of 154 \kms ; Muzerolle et al.~\cite{Mea03})
 and two non accretors { (V927~Tau and USco~CTI-O75)} out of the 37 for which there are \Macc\ 
upper limits
(Muzerolle et al.~\cite{Mea03} and this paper).
Furthermore, Fig.~\ref{haw}  shows  that there is a rather good
correlation between the \Ha\ 10\% width and \Macc\ over the whole
range of mass from BDs to TTS, so that
it is possible to use
the  observed width not only to discriminate between accretors and non-accretors but also to get an approximate
estimate of the accretion rate, without performing detailed
model fits.
We show in Fig.~\ref{haw} the best-fit relation between these two
quantities for \Ha\ 10\% width $>200$ \kms , which can be
expressed as:

\begin{equation}
{\rm Log} \dot M_{ac} = -12.89 (\pm 0.3) + 9.7 (\pm 0.7)\times 10^{-3} \,\rm H\alpha 10\%
\end{equation}
 where $\rm H\alpha 10$\% is the \Ha\ 10\% width in \kms\ and \Macc\ is in \Myr.

The spread is rather large, and can be due in part to the fact that
in most cases the measurements used to derive \Macc\ and the high-resolution
\Ha\ profiles have not been obtained simultaneously. We show, as an
illustration of possible problems, the rather unusual
case of the TTS DF~Tau 
which has an accretion rate of about $10^{-7}$ \Myr, based on 
veiling from medium-resolution
spectrophotometric data obtained in 1996 (Gullbring et al.~\cite{Gea98}).
High resolution profiles of several Balmer lines
obtained non-simultaneously  from 1988 to 1990
by Edwards et al.~(\cite{Eea94}), show broad emission in all the
lines, but with large discrepancies between them; in particular, the
\Ha\ width is 
smaller than that of the higher Balmer lines.
One could certainly improve the correlation  if  larger and simultaneous
sets of data for TTS were available.
\Macc\ values derived in this way are necessarily inaccurate for individual 
objects, and should be used with
care. Nevertheless,
they  can be very useful when dealing with large samples of objects.


\begin{figure}
\begin{center}
\leavevmode
\centerline{ \psfig{file=ps.haw,width=9cm,angle=0} }
\end{center}
\caption{ Accretion rate as a function of
the \Ha\ full width at 10\% peak intensity.
Dots (full and open) and pentagons  are
objects  with mass $\simless 0.23$ \Msun. Full dots are
from this paper, open dots from Muzerolle
et al.~(\cite{Mea03}); for  both samples 
\Macc\ has been derived from model-fits
of the \Ha\ profiles. Pentagons are  substellar objects for which \Macc\ has
been obtained from veiling measurements
(Muzerolle et al.~\cite{Mea03}, \cite{Mea00},
White and Basri \cite{WB03}; Barrado y Navascu\'es et al.~\cite{BMJ04}). 
The small dots
at Log~\Macc=$-12$
are VLMOs with no evidence of accretion
(37 in total); the value of the accretion rate is the model-determined
upper limit. 
Squares are TTS (mass $\simgreat$ 0.3 \Msun)
for which accretion rates have been measured from veiling. 
Most are TTS in Taurus-Auriga (\Macc\ from
Gullbring et al.~\cite{Gea98}, \Ha\ widths 
from Edwards et al.~\cite{Eea94}). We have also added TW Hya 
(Muzerolle et al.~\cite{Mea00}) and GN~Tau from White and Basri (\cite{WB03}).
The dashed line shows the best-fitting relation
(Eq.~1).
The four  connected squares show the 10\% full width of \Ha\, H$\beta$,
H$\gamma$ and H$\delta$ for DF Tau from Edwards et al.~(\cite{Eea94});
\Ha\ is the narrowest of the four.
}
\label{haw}
\end{figure}

\subsection {\Pab\ and \Brg\ as accretion indicators}

\begin{figure}
\begin{center}
\leavevmode
\centerline{ \psfig{file=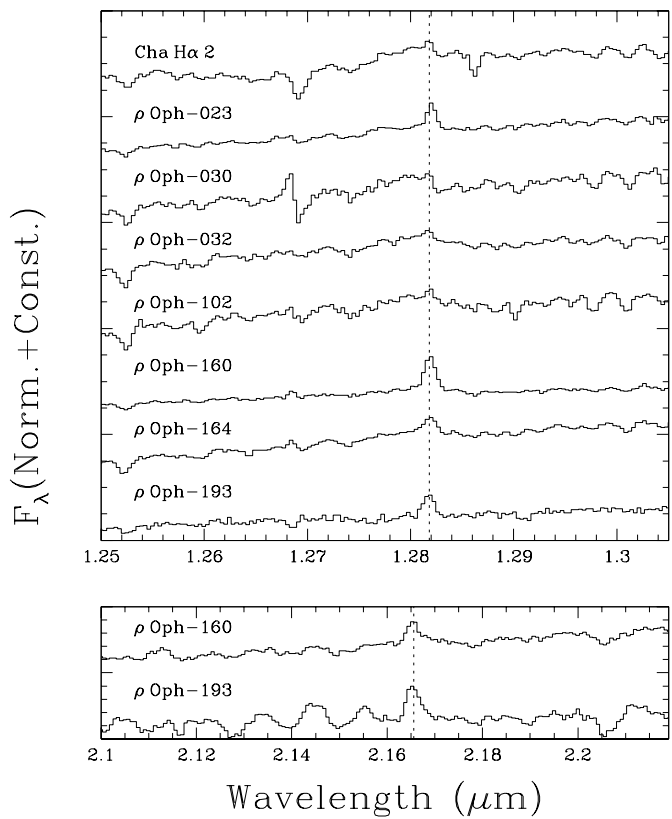,width=9cm,angle=0} }
\end{center}
\caption{ Near-IR spectra of objects with \Pab\ (top panel)
or \Brg\ (bottom panel) detection.}
\label{ir_spectra}
\end{figure}

Our near-IR spectroscopy does not allow us to resolve the line
profiles, and we can only measure
equivalent widths, listed in Table~1.
Of the 9 Chamaeleon I objects,
\Pab\ is detected in only 1. For the others, we can set
upper limits to the equivalent width (3$\sigma$) of about 0.3--0.9 \AA.
In Ophiucus, on the contrary, we detect \Pab\ emission in 7 of the
 10 observed objects.
\Brg\ is detected only in 2 objects in  \Roph.
Of the 5 objects with evidence of accretion from \Ha , \Pab\ is detected
in 4, while \Brg\ is not detected in any.
{Fig.~\ref{ir_spectra} shows the spectra of the objects with line detections.}


{ 
These results suggest that \Brg, which is intrinsically weaker than 
\Pab, is more difficult to detect even in
highly reddened objects, where  differential extinction favors lines
at longer wavelengths.} For this reason,
we will focus in the following on the use of \Pab\ as an accretion indicator.
We have therefore converted equivalent widths into fluxes using
broad-band magnitudes and then correcting for extinction
for the  Cha~I objects, while
for the \Roph\ objects we have used the
calibrated low-resolution spectra of Natta et al.~(\cite{Nea02}) for the 
continuum and 
corrected for extinction the resulting line fluxes.

\begin{figure}
\begin{center}
\leavevmode
\centerline{ \psfig{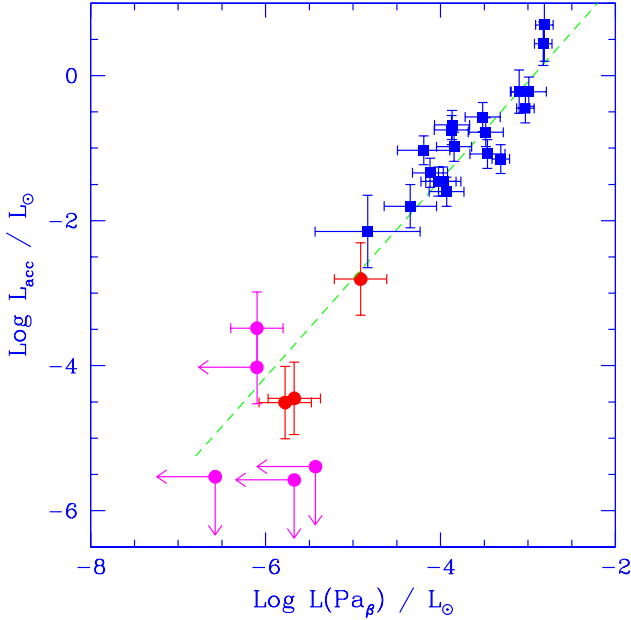} }
\end{center}
\caption{ \Pab\ luminosity as a function of \Lacc. The dots are
the VLMOs from this paper. The squares are T Tauri stars  in Taurus-Auriga from
Muzerolle et al.~(\cite{Mea98}). The dashed line is the best-fitting
relation (Eq.~2).}
\label{lab}
\end{figure}

Fig.~\ref{lab} shows the relation between the \Pab\ luminosity (derived from
the measured fluxes assuming a distance of 150 pc for Ophiucus and 160 pc for
Cha~I) and the accretion luminosity, computed from \Macc\ and the stellar
parameters given in Table~1. According to Muzerolle et al.~(\cite{Mea98}),
these  two quantities (rather than line flux versus
\Macc) show the tightest correlation in TTS, and we have
added Muzerolle's sample to ours.
The figure shows that the trend of lower
\Pab\ luminosity for lower \Lacc\ extends to the very low values of \Lacc 
which characterize VLMOs.
In fact, one can derive a rather good relation
to estimate \Lacc\ from the \Pab\ luminosity over the whole
range of masses from few tens of Jupiter masses to about one
solar mass, shown by the dashed line in Fig.~\ref{lab} and
given by:

\begin{equation} 
{\rm Log} L_{ac} = 1.36 (\pm 0.2) \times {\rm Log} L(Pa_\beta) + 4.00 (\pm 0.2)
\end{equation}
where \Lacc\ and L(\Pab) are in units of \Lsun. 
Note that this relation is, within the uncertainties, identical to that
derived by Muzerolle et al.~(\cite{Mea98}) for  TTS 
over a much narrower interval of L(\Pab).

As for \Ha\, we have ignored the possibility that 
some \Pab\ emission may come from wind or jets, driven by  accretion.
In several TTS, the profiles of \Pab\ and \Brg\ are well described
by magnetospheric accretion models (Folha and Emerson \cite{FE01}),
but there are
some examples of  \Pab\ wind emission 
(Whelan et al.~\cite{Whea04}).
Inspection of Fig.~\ref{lab} shows that, { very likely, any correction for
a wind/jet contribution to the \Pab\ luminosity will be lost
within the uncertainties of the individual measurements
and the large  scatter of the points.}
Nevertheless, the possibility of
a contribution  from outflowing matter to lines should be kept in mind when
discussing individual objects.

Using the relation (2), we have derived \Lacc\ and \Macc\ for all the objects
in our sample; \Macc\ is given in the last column of Table~1.
The largest differences between this determination of \Macc\ and that from
the \Ha\ profile is of the order $\pm$0.7--0.8 in log, i.e., a factor $\sim$6
(see also Fig.~\ref{lab}).

\subsection {Variability}

Spectroscopic and photometric variability over a large interval of timescales
is typical of pre-main sequence stars of all mass, including sub-stellar
objects. By necessity, this introduces an artificial spread
in correlation such as those of Fig.~\ref{haw} and \ref{lab}, when
the accretion rate or luminosity and the line properties are not
measured simultaneously.
We expect that the quality of these correlations could improve significantly
if simultaneous observations were available.

Another consequence of variability
is the difficulty of assigning a well-defined accretion rate to any 
specific star. 
Variations of the \Brg\  flux are typically within a factor
2--3 (see, for example, Greene and Lada \cite{GL97};
Luhman and Rieke \cite{LR98});  if similar variations apply to \Pab\
(for which we could not find multiple-epoch data in the literature)
this would  change the estimate of \Lacc\
by a factor $\sim 2-5$. In some cases, the lines become undetectable, and one
would then classify the star among non-accretors.

\begin{figure}
\begin{center}
\leavevmode
\centerline{ \psfig{file=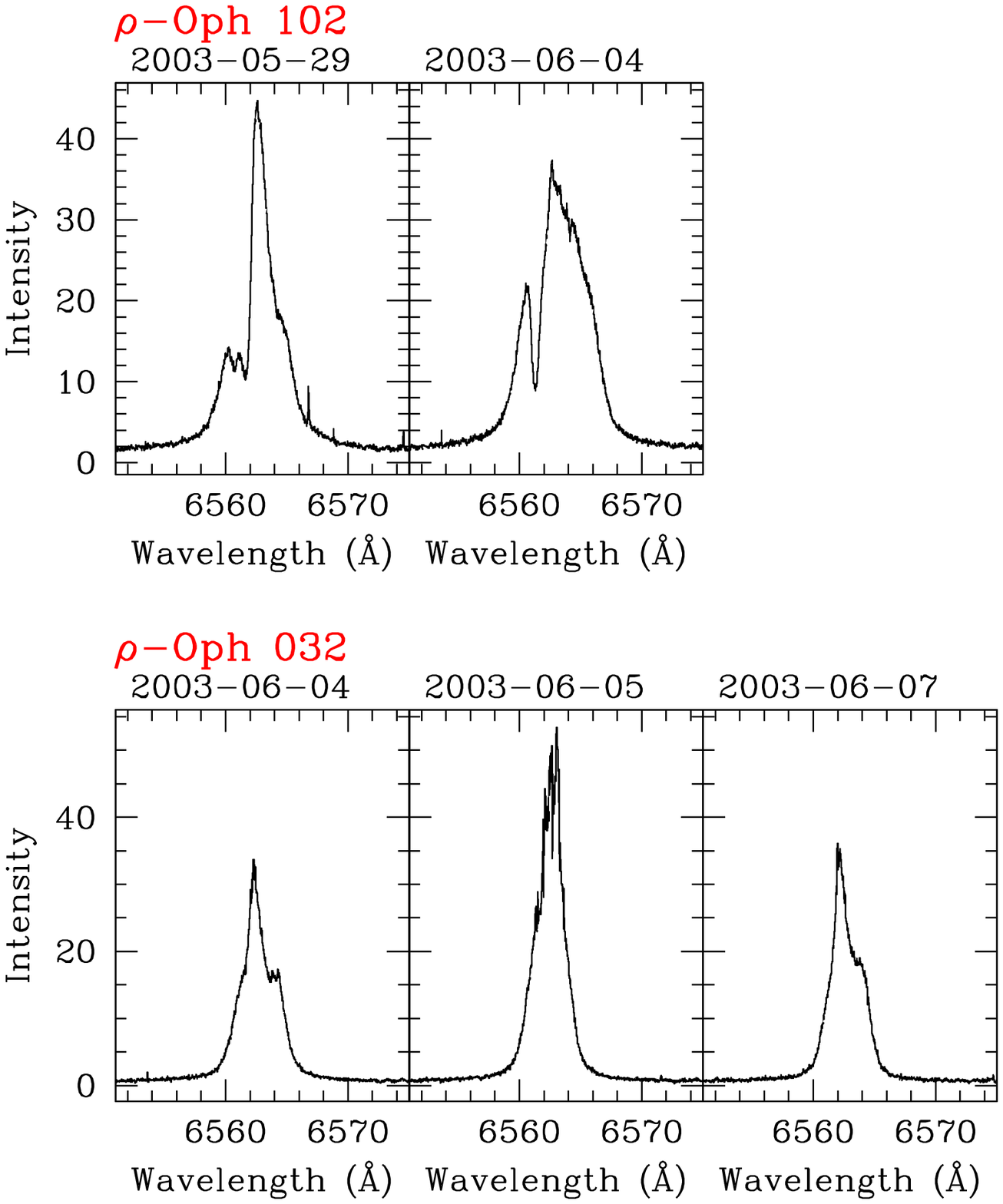,width=8.5cm,angle=0} }
\end{center}
\caption{\Ha\ variability in \Roph -102 and \Roph -032.}
\label{Ha_ind}
\end{figure}

Similar uncertainties are obtained if we consider the variations of the \Ha\ 10\%
width. In our spectra (see Fig.~\ref{Ha_ind}),
the largest variation is seen in \Roph -102, where the \Ha\ 10\% width varies
from 335 \kms\ on May 29, 2003 to 411 \kms\ on June 4, 2003. According
to eq.(1), the corresponding accretion rate would increase by a factor $\sim$5.
In all the other objects the \Ha\ variations are smaller, implying a
maximum change of \Macc\ of a factor $\sim$3 at most.
A comparison of the sub-stellar objects in IC 348
for which \Ha\ profiles have been obtained by Muzerolle et al.~(\cite{Mea03})
and Jayawardana et al.~(\cite{Jay03b}) shows variations that would results
in changes of \Lacc\ again by a factor $\sim 2-5$, according to Eq.(1).

An interesting case is that of \Roph -030 (GY~5),  which has a strong
\Ha\ emission in our and Jayawardana et al.~(\cite{Jay02}) spectra
(but a very different
profile and a 10\% \Ha\ width of 292 and 352 \kms\, respectively) 
but is narrow (177 \kms)  and weak in the spectrum of
Muzerolle et al.~(\cite{Mea03}). 

Some of these variations  may be
due to differences in the quality of the spectra (resolution, signal-to-noise
etc.), since all the  VLMOs
are very weak, often close to
the instrumental limits. Still, most of the variability is likely to be real,
as suggested by very similar results in TTS. For any individual object,
one should have 
high quality spectra and long time sequences, which could 
provide not an instantaneous but a ``typical" accretion rate.
However, if we consider a large sample of objects, these individual
variations will average up
and will only increase  the spread of any existing
correlation, but not change it  systematically.

\subsection {Other Optical Lines}

The UVES spectral range contains a number of other emission
lines (in addition to \Ha),
usually considered  good indicators of mass-loss and accretion.
Table \ref{other_lines} gives the equivalent width of the lines (positive values
for emission) and, in some cases, as for the NaD doublet, some information
on their profile. The spectrum of Cha \Ha 1 is too noisy, and we have
dropped this object from the table.

The HeI lines are generally interpreted as emitted in
the accretion
shock at the base of the infalling matter: these lines
are indeed present and measurable in 
all the objects where we detect accretion in \Ha\ (Cha \Ha 2,
Cha \Ha 6, \Roph -030, \Roph -032 and \Roph -102), but not
in Cha \Ha 3 and 5, which have narrow \Ha\ profiles and no
evidence of accretion. This result is consistent with our
\Ha\ analysis.

\setcounter{table}{2}
\begin{table*}
\begin{flushleft}
\caption{ Equivalent width of other optical emission lines}
\begin{tabular}{lcccccccc}
\hline\hline
&&&&&&&&\\
Object  & [OI]$\lambda$6300& [OI]$\lambda$6363&  [NII]$\lambda$6583& [SII]$\lambda$6716& [SII]$\lambda$6731& HeI$\lambda$5875& HeI$\lambda$6678&NaI$\lambda$5896,5890$^a$\\
        & (\AA)& (\AA)& (\AA)& (\AA)& (\AA)& (\AA)& (\AA)&\\
&&&&&&&&\\
\hline
Cha \Ha 2& -- & --  & 0.06:& 0.06:& $<$0.1& 1.6& 0.6& E(PCyg)\\
Cha \Ha 3& -- & --  & 0.03:& $<0.1$ & 0.08 & $<0.1$ & $<0.1$ & E \\
Cha \Ha 5& -- & --  & $ <0.1$ & $<0.08$ & $<0.1$ &$<$0.2& $<0.08$ & E\\
\medskip
Cha \Ha 6&--  & --  & 0.07 & $<0.08$ & $<$0.08 & 1.1& 0.3& E\\
\Roph -030& --  &--& $<0.1$ & $<0.08$ & $<0.08$ & 1.0& 0.3: & E\\
\Roph -032& -- & -- & 0.1: & 0.12: & $< 0.1$& 6.0& 2.2& E\\
\Roph -102& 2:& 0.4: & 0.1: & 0.07& 0.2:& 4.4& 2.3& E(PCyg)\\
\hline
\hline
\label{other_lines}
\end{tabular}
\end{flushleft}
$^a$  E: both components in emission; PCyg: additional blue-shifted
absorption (see Fig.~\ref{NaD}).
\end{table*}

\begin{figure}
\begin{center}
\leavevmode
\centerline{ \psfig{file=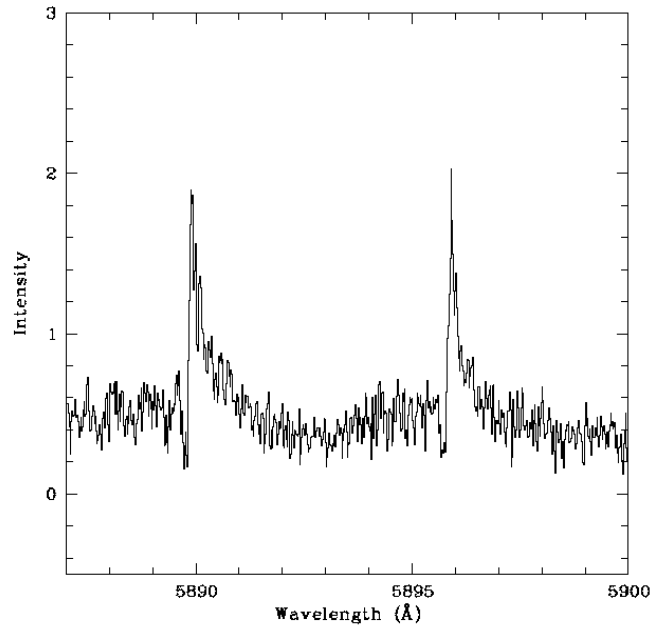,width=9cm,angle=0} }
\end{center}
\caption{NaD lines in \Roph -102.}
\label{NaD}
\end{figure}

Some of the objects show indications of outflows. These are particularly
strong in \Roph -102, which has emission in  all the forbidden lines
of [OI], [NII], [SII], and shows a P-Cygni profile in the
NaD doublet (see Fig.~\ref{NaD}), but also \Roph -32, Cha \Ha 2 and Cha \Ha 3
have detectable
emission in some of the forbidden lines, and clear emission
in the NaD doublet. All these objects are good candidates for
further studies of winds and jets in VLMOs.
However, there is no evidence of H$_2$ vibrationally
excited emission in the IR spectra of our objects, with the possible
exception of \Roph-033, where, however, it is likely due to contamination
from VLA~1623 (see also Williams et al.~\cite{Wea95}).

Weak accretion-driven winds are expected in VLMOs, if they behave as TTS.
Note, however,
that even in \Roph -102 the equivalent width of \Ha\ is much
larger than that of the [SII] lines (the ratio is $\sim$ 200). 
The weakness of the forbidden lines in comparison to \Ha\ 
in our objects supports the assumption that the hydrogen lines form mostly in
the accreting gas (see also
Fern\'andez and Comer\'on \cite{FC01} and references
therein). 

As well known the presence of Li is a strong indication of substellar nature
and/or youth.
The covered spectral range includes the Li~{\sc i} doublet at 6708~\AA. 
We detect it in absorption in all but one objects with equivalent widths ranging
between 0.2 and 0.5~\AA, confirming, in combination
with the late-M spectral types,  that they are
members of the clouds. The Li line
is not detected in Cha \Ha 1; we believe this is due to the very
low S/N of the spectrum which shows virtually no continuum
(see also J\"orgens and G\"unther \cite{JG01}).

\section{Discussion}

\subsection {Accretion as a function of mass and age}

Our sample of VLMOs more than double the number
of objects with measured  \Macc.  
We plot in
Fig.~\ref{acc} our results, together
with  a compilation of  data  from the literature,
as a function of the mass of the central object. The data
include
TTS  and VLMOs, spanning a range of masses of
about two orders of magnitude.
All the masses have been redetermined self-consistently using the
same evolutionary tracks (1998 unpublished update
of D'Antona and Mazzitelli \cite{DM97}),
and the accretion rates have been scaled to these values, when necessary.
Note that we have plotted upper limits for \Macc\ only for 
objects from this paper.

Both \Macc\ and \Mstar\ have large uncertainties, discussed in detail,
among others,
by Muzerolle et al.~(\cite{Mea03}) and in this paper.
There are, however, some clear trends.
First of all, Fig.~\ref{acc} confirms, over a large range of masses, 
the
trend already found by other authors
(Muzerolle et al.~\cite{Mea03}; Rebull et al.~\cite{Rea00}) of
decreasing \Macc\ for decreasing \Mstar.
If, for example,  we compare the median \Macc\ for \Mstar$<$0.1 \Msun\
to that in the mass interval 1--0.3 \Msun, we find values of
$\sim 3\times 10^{-10}$ and $\sim 10^{-8}$ \Myr, respectively (neglecting upper limits to \Macc).
It is
unlikely that this trend is caused by sensitivity
limits. Although, as we will discuss in the following section,
in many VLMOs the accretion rate is close to the model sensitivity,
this is not the case of TTS (see, e.g., Muzerolle et al.~\cite{Mea03}),
so that the correlation we find neglecting upper limits is probably
somewhat shallower than the true one.

A second effect is  clearly shown in Fig.~\ref{acc}, namely that
the accretion rate, for the same mass of the central object, is higher
in younger regions. The \Roph\ brown dwarfs have accretion rates
of roughly $10^{-9}$--$10^{-10}$ \Myr, more than an order of
magnitude larger than most objects of similar mass in Taurus, Chamaeleon
and IC~348. 
An age dependence of \Macc\ is known in TTS (e.g., Calvet et al.~\cite{CHS00}),
and is suggested for VLMOs by the fact
that the fraction of objects with
narrow \Ha\ profiles (non-accretors)
is larger in older star forming regions (Jayawardana et al.~\cite{Jay02}, 
 \cite{Jay03b}). Our
results confirm this trend:  only 10\% (1 out of 10) of the
 objects in  our
Cha~I sample are accreting; if we include also the G\'omez and Persi
(\cite{GP01}) VLMOs, then the fraction becomes 28\% (5 out of 18 objects). 

The dependence of \Macc\ on age
makes it hard to determine the 
exact relationship of \Macc\ vs. \Mstar\ from samples where objects
in regions of very different age are collected together, as 
in Fig.~\ref{acc}.
For example, the \Roph\ BD population selected by Natta et al.~(\cite{Nea02})
is particularly young, and it would be interesting to compare their
accretion rates to equally young TTS in \Roph, rather than to TTS
in Taurus.
Also, there are a few VLMOs in Taurus, Cha~I and IC348 with
very large accretion rates, comparable or even higher than those of
the \Roph\ VLMOs. Some of these estimates need to be confirmed,
and the possibility of having detected a flare ruled out. Still, 
one wonders if they could be much younger than the
average population of the star-forming region to which they belong.

In spite of the large uncertainties that affect individual objects, 
these preliminary results
show that it is now possible, with the advent of 8-m class telescopes
that give access to the VLMOs population,
to quantify the dependence of \Macc\ on mass and 
age, overcoming, due to the large range of \Mstar\ that one can explore,
the  uncertainties of  individual measurements.
For this one  needs larger
and more homogeneous samples of stars (not only VLMOs but also
TTS) in a variety of star-forming regions.



\begin{figure}
\begin{center}
\leavevmode
\centerline{ \psfig{file=ps.acc,width=9cm,angle=0} }
\end{center}
\caption{Accretion rate as a function of the mass of the central object.
Filled dots and  squares are VLMOs in \Roph\ and Cha~I, respectively,
from this paper; \Macc\ has been determined from L(\Pab) as described
in \S 3.3. Arrows are 3$\sigma$ upper limits. 
Open squares are objects in Cha~I from G\'omez and Persi (\cite{GP01})
with \Pab\ emission, where \Macc\ has been determined from L(\Pab).
Objects in
IC 348 (Muzerolle et al.~\cite{Mea03}) are shown by diamonds,
in TW Hya (Muzerolle et al.~\cite{Mea00}) by pentagons,
in Taurus (White and Ghez \cite{WG01}, White and Basri \cite{WB03},
Muzerolle et al.~\cite{Mea03}) by crosses. We do not plot
objects from the literature 
with no detected accretion.
}
\label{acc}
\end{figure}

\subsection {Disks and Accretion}

Classical TTS show a clear correlation between accretion and the presence
of a circumstellar disk. Disks are detected around a number of VLMOs,
which show mid-infrared fluxes well in excess of the photospheric ones
(e.g., Comer\'on et al.~\cite{Cea98},
Persi et al.~\cite{Pea00}, Bontemps et al.~\cite{Bea01}),
modeled as the emission of 
disks heated by radiation from the central objects
(Natta and Testi \cite{NT01}, Testi et al.~\cite{Tea02};
Natta et al.~\cite{Nea02}). 

Table~\ref{disks} summarizes the information on the presence of a mid-IR
excess in our sample objects.
All  10 objects in \Roph\ have been detected by ISOCAM at 6.7 and 14.3 \um\
(Bontemps et al.~\cite{Bea01}), 
and the correlation between disk and accretion, i.e., broad \Ha\ and/or emission in the hydrogen IR lines, is basically confirmed by the results
of this paper.

\begin{table}
\begin{flushleft}
\caption{ Correlation between accretion and mid-IR excess}
\begin{tabular}{lccccc}
\hline\hline
&&&&&\\
Star  & Acc. & Acc.& Acc.& 6.7 \um & 14.3 \um\\
      & (\Ha )     &  (\Pab)  &  (\Brg)  &         &     \\
&&&&&\\
\hline
Cha \Ha 1& N & N & N& {\bf Y}& {\bf Y}\\
Cha \Ha 2& {\bf Y}& {\bf Y}& N& {\bf Y}& {\bf Y}\\
Cha \Ha 3& N & N& N& {\bf Y}& N\\
Cha \Ha 5& N& N& N&  {\bf Y}& N\\
Cha \Ha 6& {\bf Y}& N& N& {\bf Y}& N\\
Cha \Ha 7& --& N& N& N& N\\
Cha \Ha 9& --& N& N& {\bf Y}& {\bf Y}\\
Cha \Ha 10&--& N& N& N& N\\
\medskip
Cha \Ha 11& --& N& N& N& N\\
\Roph -023& --& {\bf Y}& N& {\bf Y}& {\bf Y}\\
\Roph -030& {\bf Y}& {\bf Y}& N& {\bf Y}& {\bf Y}\\
\Roph -032& {\bf Y}& {\bf Y}& N& {\bf Y}& {\bf Y}\\
\Roph -033& --& N& N& {\bf Y}& {\bf Y}\\
\Roph -102& {\bf Y}& {\bf Y}& N& {\bf Y}& {\bf Y}\\
\Roph -160& --& {\bf Y}& {\bf Y}& {\bf Y}& {\bf Y}\\
\Roph -164& --& {\bf Y}& N& {\bf Y}& {\bf Y}\\
\Roph -176& --& N& N& {\bf Y}& {\bf Y}\\
\Roph -193& --& {\bf Y}& {\bf Y}& {\bf Y}& {\bf Y}\\
\Roph -GY10& --& N& N& {\bf Y}& {\bf Y}\\
\hline
\hline
\label{disks}
\end{tabular}
\end{flushleft}
\end{table}

{ 
The Chamaeleon sample is in a way more intriguing. Three objects
(Cha \Ha 1, Cha \Ha 2, Cha \Ha 9) have excess emission at 6.7 and 14.3
\um (Persi et al.~\cite{Pea00}), modeled by Natta and Testi (\cite{NT01}) as due to disks.
Of these, Cha \Ha 2 is clearly accreting, Cha \Ha 1  and 
Cha \Ha 9 (for which we have no \Ha\ profile) do not.
Three objects have been detected at 6.7 \um,
but not at 14.3 \um; of these, Cha \Ha 3 and Cha \Ha 5 do not show
evidence of accretion in \Ha\ nor in \Pab\ or \Brg,
while Cha \Ha 6 has broad \Ha, but no emission in the IR lines.
Finally, three (\Ha 7, \Ha 10 and  \Ha 11) have no detected mid-IR excess 
and no IR line emission.
}

It appears from these results that disks are detected more easily than
accretion, and that there is a population of VLMOs with evidence of disks but no detectable accretion activity. This is somewhat different from TTS, where
most WTTS have no evidence of accretion but also no detectable IR excess
(as some of the Cha I objects).
This is probably due to the fact that even in very active  VLMOs
the accretion rate  is close
to the sensitivity limit of the various methods one can use, making it difficult to discriminate between ``accretors'' and ``non-accretors''.
On the contrary, in TTS
typical accretion rates are
at least two orders of magnitude higher than the sensitivity limit
of veiling measurements and/or \Ha\ profile fitting (Gullbring et al.~\cite{Gea98};
Muzerolle et al.~\cite{Mea03}), and the  separation between accreting and non-accreting objects is therefore much more meaningful.

One of the implications of the very low accretion rates that characterize
BDs is that their disks should live long. If we assume, for
example, that the disk mass scales with the mass of the central objects
in BDs as in TTS, we have typical disk masses of about $10^{-3}$ \Msun;
with an accretion rate of $\sim 3\times 10^{-10}$ \Myr, as
measured in \Roph, the disk will live at least $3\times 10^6$ years, as a typical
TTS. This agrees with the results of Liu et al.~(\cite{Liu03}), who
found no signficant difference in the fraction of disks between BDs and
TTS, and that the age of the TTS with disks was similar to that of TTS
without disks. 

It is more difficult to use this same result to rule out ``truncated" disks,
as those produced by the dynamical ejection of stellar embryos
predicted, e.g., by Bate et al.~(\cite{BBB03}). { Let us assume, for example,
that the disk has an outer radius of 100 AU, and that, once ejected, 
loses all  material outside a  radius of about 10 AU.
The mass left depends critically on the unknown density profile of the
disk, and may range from about 1/3 (if the surface density profile
was $\Sigma \propto R^{-1.5}$) to about 1/10 (for $\Sigma \propto R^{-1.0}$)
of the original mass.}
A disk with mass of only $10^{-4}$ \Msun will  live about $3\times 10^5$
years,
much less than a typical disk of TTSs, {
and also much less than the age of many
BDs with infrared excess in star-forming regions (e.g., Jayawardhana et 
al.~\cite{Jay03a}).}
However, if the density profile
is very peaked, the difference will not be significant, given the large
uncertainties and spread of all the parameters involved. Only very accurate
model predictions can be verified or falsified by this kind of
arguments.

A final point worth  mentioning is that the relation of the accretion rate
with the mass and age
of the central object may shed light on the physical process
by which disk matter accretes onto the central star.
The simplest suggestion is that the accretion rate {may be related to the temperatures
in the disk, which in turn are closely coupled to the luminosity of the
central object.  This may explain both the  dependence of \Macc\
on \Mstar\ and its decrease with time, since objects of all masses, and
BDs in particular, fade as they get older. However, it is not clear that the change in temperature is big enough to account for the changes in \Macc.}
More subtle effects can be
at work, related for example to the level of chromospheric and
coronal activity
(and hence X-ray fluxes) of the star
and resulting disk ionization (see discussion in Muzerolle et al.~\cite{Mea03}). 

\section {Summary and conclusions}

We present in this paper new optical and near-infrared spectroscopic observations of a sample of 19 very low mass objects in the regions Chamaeleon I
and \Roph. Our main aim was to test if accretion rates could be determined
from the intensity of the hydrogen recombination lines in the infrared,
namely \Pab\ and \Brg.
This possibility is very important when dealing with large
numbers of VLMOs in young star-forming regions such as \Roph, where
high-resolution spectroscopy in the visual is possible
for only a handful of objects.

To this purpose, we have obtained high-resolution \Ha\ profiles of all the
objects in our sample bright enough to be detectable with UVES on VLT.
The comparison of the observed profiles with those predicted by magnetospheric
accretion models (Muzerolle et al.~\cite{Mea01}) shows that five objects
are actively accreting,
with \Macc\ in the range $\sim 10^{-9} - 10^{-11}$ \Myr, while the remaining three
in Cha~I have narrow and symmetric profiles, and \Macc$\simless 10^{-12}$ \Myr.

\Pab\ is detected in emission in 7 of the 10 \Roph\ objects, but only in one in Cha~I.
We show that the correlation between the \Pab\ luminosity
and the accretion luminosity \Lacc, found by Muzerolle et al.~(\cite{Mea98})
for TTS in Taurus over a range of masses $\sim$0.3--1 \Msun, extends
to  masses about ten times lower.
Whereas the relation between L(\Pab) and \Lacc\ (eq.(2)) can certainly be
improved when more data will be available, it is already
reasonably well defined, and can  be used to measure \Lacc,
and consequently \Macc, for large samples of obscured VLMOs.
The results are so far less conclusive for \Brg\, for which we have a detection
in only 2 objects, none with measured accretion rate.

Using eq.(2) and the measured
L(\Pab), we determined \Macc\ for all the VLMOs in our sample,
increasing the number of VLMOs with known \Macc\ by more than a factor of two.
When plotted as a function of the mass of the central object
together with all the existing data on VLMOs and TTS from the literature,
our results confirm the trend of lower \Macc\ for lower \Mstar,
although with a large spread. Some of the spread is likely due to
an ``age" effect, namely that, for a given value of \Mstar, \Macc\
decreases strongly with time. Our very young VLMOs in \Roph\ have
on average \Macc\ at least one order of magnitude higher than objects
in older star forming regions.
The dependence of \Macc\ on central mass and time needs to be better
constrained by future observations. However, it is
already clear that these data can provide very valuable
information on the  accretion mechanism in disks during the pre-main--sequence
evolutionary phase.

One side product of our analysis is that the width of \Ha\ measured at
10\% peak intensity correlates quantitatively with \Macc\ over a large
range of \Macc\ and \Mstar. This supports the currently used criterion
to separate accretors from non-accretors among VLMOs according
to the \Ha\ 10\% width (White and Basri \cite{WB03})
and confirms that a value of 200 \kms\ {(as suggested by Jayawardana 
et al.~\cite{Jay03b})} is a rather
good threshold value. Furthermore, it suggests that one can
use the \Ha\ 10\% width to obtain a quantitative estimate of \Macc\ over
a large range of masses. Although the spread of the correlation
(eq.(1)) is large, and there are a number of
caveats to keep in mind, the possibility of  estimating 
\Macc\ without going
through rather time consuming model fits is interesting, and deserves further
attention.

Finally, we would like to note that high-resolution near-infrared 
spectroscopy on 8-m class telescopes is becoming available.
As with \Ha\ for visible objects,
high quality infrared line profiles can be compared to
 model predictions  to derive
accretion rates for obscured ones.
The additional
advantage of the IR lines is that they are likely less contaminated
than \Ha\ by wind/jet emission, flares and stellar activity
in general;  a sufficiently large  sample of objects
with measured  \Ha\ and \Pab\ profiles
will be very valuable in providing more "calibration" 
points for the line intensity--accretion rate relations, and in better
constraining   magnetospheric accretion models.

\begin{acknowledgements}
This work was partly supported by  the MIUR-cofin
grant  "Low-mass stars and brown dwarfs: formation mechanisms, mass distribution and activity".
We thank the referee (S. Mohanty) for comments that have greatly improved our paper.
\end{acknowledgements}

%
%


\end{document}